\documentclass[useAMS,usenatbib]{mn2e}
\usepackage{graphicx}
\usepackage{amssymb}

\title[] {On the spectrum of the pulsed gamma-ray emission from 10MeV to 400GeV of the Crab pulsar.}

\author[]{N. Chkheidze$^{1}$,\thanks{E-mail:
nino.chkheidze@iliauni.edu.ge} G. Machabeli$^{1}$ and Z. Osmanov$^{2}$ \\
$^{1}$Centre for Theoretical Astrophysics, ITP, Ilia State University, Tbilisi, 0162, Georgia\\
$^{2}$Free University of Tbilisi, Tbilisi, 0183, Georgia}
\begin{document}

\pagerange{\pageref{firstpage}--\pageref{lastpage}} \pubyear{2011}

\maketitle

\label{firstpage}

\begin{abstract}
In the present paper a self-consistent theory, interpreting the
VERITAS and the MAGIC observations of the very high energy pulsed
emission from the Crab pulsar is considered. The photon spectrum
between 10MeV and 400GeV can be described by two power-law functions
with the spectral indexes equal to 2 and 3.8. The source of the
pulsed emission above 10MeV is assumed to be the synchrotron
radiation, which is generated near the light cylinder during the
quasi-linear stage of the cyclotron instability. The emitting
particles are the primary beam electrons with the Lorentz factors up
to $10^{9}$. Such high energies by beam particles is supposed to be
reached due to Landau damping of the Langmuir waves in the light
cylinder region. This mechanism provides simultaneous generation of
low (radio) and high energy (10MeV-400GeV) emission on the light
cylinder scales, in one location of the pulsar magnetosphere.
\end{abstract}


\section{Introduction}

The recent observations of the Crab pulsar in the very high energy
(VHE) domain with the VERITAS array of atmospheric Cherenkov
telescopes revealed pulsed $\gamma$-rays above $100$GeV
\citep{veritas}, which was later confirmed by measurments of the
MAGIC Cherenkov telescope \citep{alek1,alek2}. Prior to the work
\citet{veritas} the highest energy of the pulsed emission from the
Crab pulsar was $25$GeV \citep{magic}. The detection of such a high
energy pulsed $\gamma$-rays cannot be explained on the basis of
current pulsar emission models. It is generally assumed that the VHE
emission is produced either by the Inverse Compton scattering or by
the curvature radiation. By analyzing the aforementioned emission
processes \citep{difus,difus1}, we have found that for Crab pulsar's
magnetospheric parameters even very energetic electrons are unable
to produce the photon energies of the order of $25$GeV. Studying the
curvature radiation, it was shown that the curvature drift
instability efficiently rectifies the magnetic field lines, leading
to a negligible role of the curvature emission process in the
observed VHE domain \citep{forcefree}. In previous work \citet{ch10}
we have explained the origin and the measured spectrum of the Crab
pulsar in the high energy (HE) domain $(0.01-25)$GeV, relying on the
pulsar emission model first developed by \citet{machus1}. According
to these works, in the electron-positron plasma of a pulsar
magnetosphere the low frequency cyclotron modes, on the quasi-linear
evolution stage create conditions for generation of the HE
synchrotron radiation.

It is well known that the distribution function of relativistic
particles is one dimensional at the pulsar surface, because any
transverse momenta ($p_{\perp}$) of relativistic electrons are lost
in a very short time (\(\leq10^{-20}\)s) via synchrotron emission in
very strong magnetic fields. But plasma with an anisotropic
one-dimensional distribution function is unstable which inevitably
leads to the wave excitation process. The main mechanism of the wave
generation in plasmas of the pulsar magnetosphere is the cyclotron
instability, which develops at the light cylinder length-scales (a
hypothetical zone, where the linear velocity of rigid rotation
exactly equals the speed of light). During the quasi-linear stage of
the instability a diffusion of particles arises as along, also
across the magnetic field lines. Therefore, plasma particles acquire
transverse momenta and, as a result, start to radiate in the
synchrotron regime.

In \citet{ch10}, it was shown that near the light cylinder the radio
waves are generated, provoking the re-creation of the pitch angles
and the subsequent synchrotron radiation in the HE domain. Thus, in
the framework of this model generation of low and high frequency
waves is a simultaneous process and it takes place in one location
of the pulsar magnetosphere. This explains the observed coincidence
of the HE emission pulses with the radio signals. The recent VERITAS
and MAGIC observations have shown that the aforementioned
coincidence takes places in the VHE domain ($>100$GeV) as well
\citep{magic,veritas,alek1}. According to \citet{veritas} detection
of pulsed $\gamma$-ray emission of the order of $100$GeV requires
that the emission should be produced far out in the magnetosphere.
Thus, we suppose that the pulsed high and the very high energy
radiation of the Crab pulsar is generated through the synchrotron
mechanism at the light cylinder length-scales, switched on due to
the quasi-linear diffusion. The resonant particles are the primary
beam electrons with the Lorentz-factor $\gamma_{b}\sim10^{8-9}$,
giving the synchrotron emission in the ($0.01-400$)GeV energy
domain.

According to \citet{magic} a joint fit to the EGRET ($10$MeV to
$10$GeV) and MAGIC ($>25$GeV) data predicted a power-law spectrum
with a generalized exponential shape for the cutoff, described as
$F_{\epsilon}\propto\epsilon^{-\alpha}\exp(-(\epsilon/\epsilon_{0})^{\beta})$,
where $\alpha=2.022\pm0.014$. We provided a theoretical confirmation
of the measured spectrum, which yielded $\beta=1.6$ and the cutoff
energy $\epsilon_{0}=23$GeV \citep{ch10}. Recent VERITAS
observations $(100-400)$GeV combined with the Fermi-LAT data
$(0.1-10)$GeV favor a broken power law as a parametrization of the
spectral shape. The good fit results are also obtained if one uses a
log-parabola function, but it fails to describe the spectrum below
$500$MeV. Although, the Fermi-LAT and Magic data below 60 GeV can be
equally well parameterized by broken power law and exponential
cutoff. In the energy range between $100$GeV and $400$GeV measured
by VERITAS and MAGIC, the spectrum is well described by a simple
power law with the spectral index equal to $3.8$ \citep{veritas,
alek2}.

The paper is organized as follows. In Sect. 2 we describe the
emission model, in Sect. 3 we derive the theoretical synchrotron
spectrum for the high and the very high energy $\gamma$-ray emission
of the Crab pulsar and in Sect. 4 we discuss our results.

\section{Emission model}

Any well known theory of pulsar emission suggests that, the observed
radiation is generated due to processes taking place in the
electron-positron plasma. It is generally assumed that the pulsar
magnetosphere is filled by dense relativistic electron-positron
plasma with  an anisotropic one-dimensional distribution function
(see Fig. 1 from \cite{arons}) and consists of the following
components: a bulk of plasma with an average Lorentz-factor
$\gamma\sim\gamma_{p}$, a tail on the distribution function with
$\gamma\sim\gamma_{t}$, and the primary beam with
$\gamma\sim\gamma_{b}$. The distribution function is one-dimensional
and anisotropic and plasma becomes unstable, which might cause a
wave excitation in the pulsar magnetosphere. The main mechanism of
wave generation in plasmas of the pulsar magnetosphere is the
cyclotron instability. The cyclotron resonance condition can be
written as \citep{kmm}:
\begin{equation}\label{1}
    \omega-k_{_{\|}}V_{_{\|}}-k_xu_x+\frac{\omega_{B}}{\gamma_{r}}=0,
\end{equation}
where $u_{x}=cV_{_{\|}}\gamma_{r}/\rho\omega_{B}$ is the drift
velocity of the particles due to curvature of the field lines with
the curvature radius $\rho$, $\omega_{B}\equiv eB/mc$ is the
cyclotron frequency, $e$ and $m$ are the electron's charge and the
rest mass, $c$ is the speed of light, $k_x$ is the wave vector's
component along the drift and $B$ is the magnetic field induction.
During the wave generation process, one also has a simultaneous
feedback of these waves on the resonant electrons \citep{vvs}. This
mechanism is described by the quasi-linear diffusion (QLD), leading
to a diffusion of particles as along as across the magnetic field
lines. Therefore, resonant particles acquire transverse momenta
(pitch angles) and, as a result, start to radiate through the
synchrotron mechanism.

The wave excitation leads to redistribution process of the resonant
particles via the QLD. The kinetic equation for the distribution
function of the resonant electrons can be written as \citep{ch10}:
\begin{eqnarray}\label{2}
\frac{\partial\textit{f }^{0}}{\partial
    t}+\frac{\partial}{\partial
p_{\parallel}}\left\{F_{\parallel}\textit{f
}^{0}\right\}+\frac{1}{p_{\perp}}\frac{\partial}{\partial
p_{\perp}}\left\{p_{\perp}F_{\perp}\textit{f }^{0}\right\}=\nonumber
\\=\frac{1}{p_{\perp}}\frac{\partial}{\partial p_{\perp}}\left\{p_{\perp}D_{\perp,\perp}\frac{\partial \textit{f }^{0}}{\partial p_{\perp}}\right\}.
\end{eqnarray}
where
\begin{equation}\label{4}
    F_{\perp}=-\alpha_{s}\frac{p_{\perp}}{p_{\parallel}}\left(1+\frac{p_{\perp}^{2}}{m^{2}c^{2}}\right),\qquad
    F_{\parallel}=-\frac{\alpha_{s}}{m^{2}c^{2}}p_{\perp}^{2},
\end{equation}
are the transversal and longitudinal components of the synchrotron
radiation reaction force, where
$\alpha_{s}=2e^{2}\omega_{B}^{2}/3c^{2}$ and $D_{\perp,\perp}$ is
the transverse diffusion coefficient which is defined as follows
\citep{ch10}
\begin{equation}\label{}
    D_{\perp,\perp}=\frac{\pi e^{4}
n_{p}}{8mc\omega_{B}^{2}\gamma_{p}^{3}}|E_{k}|^{2}.
\end{equation}
Here  $|E_{k}|^{2}$ is the density of electric energy in the waves
and its value can be estimated from the expression
$|E_{k}|^{2}\approx mc^{2}n_{b}\gamma_{b}c/2\omega_{c}$, where
$\omega_{c}$ is the frequency of the cyclotron waves. From Eq. (1)
it follows that
\begin{equation}\label{}
    \omega_{c}\approx\frac{\omega_{B}}{\delta\gamma_{r}},
\end{equation}
where $\delta = \omega_p^2/(4\omega_B^2\gamma_p^3)$, $\omega_p
\equiv \sqrt{4\pi n_pe^2/m}$ is the plasma frequency and $n_p$ is
the plasma density.

The transversal QLD increases the pitch-angle, whereas force
$F_{\perp}$ resists this process, leading to a stationary state
($\partial\textit{f}/\partial t=0$). The pitch-angles acquired by
resonant electrons during the process of the QLD satisfies $\psi=
p_{\perp}/p_{\parallel}\ll1$. Thus, one can assume that
$\partial/\partial p_{\perp}>>\partial/\partial p_{\parallel}$. In
this case the solution of Eq.(2) gives the distribution function of
the resonant particles by their perpendicular momenta \citep{ch10}

\begin{equation}\label{9}
    \textit{f}(p_{\perp})=C exp\left(\int
    \frac{F_{\perp}}{D_{\perp,\perp}}dp_{\perp}\right)=Ce^{-\left(\frac{p_{\perp}}{p_{\perp_{0}}}\right)^{4}},
\end{equation}
where
\begin{equation}\label{11}
     p_{\perp_{0}}\approx\frac{\pi^{1/2}}{B\gamma_{p}^{2}}\left(\frac{3m^{9}c^{11}\gamma_{b}^{5}}{32e^{6}P^{3}}\right)^{1/4}.
\end{equation}
And for the mean value of the pitch angle we find $\psi_0\approx
p_{\perp_{0}}/p_{\parallel}\simeq10^{-6}$. Synchrotron emission is
generated as the result of appearance of pitch angles.

The synchrotron emission flux of the set of electrons in the
framework of the present emission scenario is written as (see
\citet{ch10})
\begin{equation}\label{14}
    F_{\epsilon}\propto\int_{p_{_{\parallel _{min}}}}^{p_{_{\parallel _{max}}}}\textit{f}_{\parallel}(p_{\parallel})B\psi_{0}\frac{\epsilon}{\epsilon_{m}}\left[\int_{\epsilon/
    \epsilon_{m}}^{\infty}K_{5/3}(z)dz \right] dp_{\parallel}.
\end{equation}
Here $\textit{f}_{_{\parallel}}(p_{_{\parallel}})$ is the
longitudinal distribution function of electrons,
$\epsilon_{m}\approx5\cdot10^{-18}B\psi_{0}\gamma^{2}$GeV is the
photon energy of the maximum of synchrotron spectrum of a single
electron and $K_{5/3}(z)$ is the Macdonald function. After
substituting the mean value of the pitch-angle in the above
expression for $\epsilon_{m}$, we get
\begin{equation}\label{15}
    \epsilon_{m}\simeq5\cdot10^{-18}\frac{\pi^{1/2}}{\gamma_{p}^{2}}\left(\frac{3m^{5}c^{7}\gamma_{b}^{9}}{4e^{6}P^{3}}\right)^{1/4}.
\end{equation}
Accordingly, the beam electrons should have
$\gamma_{b}\simeq(6\cdot10^{8}-10^{9})$ to radiate the photons in
the energy domain $\sim(10-100)$GeV energy. The gap models provide
the Lorentz factors up to $10^{7}$, which is not enough to explain
the detected pulsed emission. Consequently, the additional particle
acceleration mechanism should be invoked to accelerate the fastest
electrons to even higher energies.

According to \citet{veritas} the observations of the Crab pulsar
indicates that the HE pulsed emission should be produced far out in
the magnetosphere. Therefore, we assume that the Langmuir waves (L)
generated via the two-stream instability at the light cylinder
length-scales undergo Landau damping on the fastest beam electrons,
which results in their effective acceleration. Excitation of L waves
through the two-stream instability in the relativistic
electron-positron plasma is considered in a series of works (see
e.g., \citet{1,2,3,4,5}), where it is assumed that the instability
develops due to overlapping of the fast and slow pair plasma
particles at distances $r\sim10^{8}$cm from the star surface. For
typical parameters of pair plasma in the pulsar magnetospehers the
growth rate of the instability is quite sufficient (see \citet{5}),
that inevitably provides existence of L waves in the vicinity of the
light cylinder zone. The phase velocity of the excited L waves is
asymptotically close to the speed of light \citep{5}. Therefore,
these waves can be only damped on the fastest electrons of the
primary beam, which velocity equals the phase-velocity of the waves
and the distribution function satisfies
$\partial\textit{f}_{b_{0}}/\partial p_{\parallel}<0$. Taking into
account the equipartition of energy among the plasma components
$n_{p}\gamma_{p}=n_{b}\gamma_{b}/2$, one can estimate the energy
density of L waves as the half of the energy density of the primary
beam particles. Through the Landau damping process the energy of the
waves is transferred to the small fraction of the beam electrons
with the highest Lorentz factors. If we assign the density of these
electrons as $n^{*}$ and the Lorentz factors as $\gamma^{*}$, and
equate the energy densities of particles and the L waves, we will
get $2n^{*}/n_{b}\approx\gamma_{b}/\gamma^{*}$. The total density of
the beam electrons is equal to the Goldreich-Julian density
$n_{b}=B/Pce\approx2\cdot10^{7}cm^{-3}$ \citep{gold} and also if we
take into account that $\gamma_{b}\sim10^{7}$ and
$\gamma^{*}\sim10^{9}$, we find $n^{*}\sim10^{5}cm^{-3}$. As we see,
if the wave energy is transferred to the fastest beam electrons,
which number is two orders of magnitude smaller than the total
number of the primary beam particles, they will gain the Lorentz
factors up to $10^{9}$. The observational fact, that the emission
flux above $25$GeV decreases should be caused by reduced number of
emitting particles with the highest Lorentz factors.

It should be mentioned that during the Landau damping process the
beam distribution function will be alongated and will form a high
energy 'tail'on the distribution function. The final shape of the
distribution function after the quasi-linear relaxation is the
plateau, and the stationary state is reached. But in our case this
might not be achieved as in the same region where the L waves are
damped the cyclotron instability is developed, which involves the
beam electrons into the cyclotron resonance process. This
complicated process causing the redistribution of the resonant
particles needs a more detailed investigation, which we plan to
perform in our future work. In the present paper, we estimate the
final shape of the distribution function which provides explanation
of the measured spectrum.

The beam particles lose their energy through the synchrotron
radiation, which sets the upper limit on the Lorentz factors that
can be achieved during their acceleration process \citep{de}. The
maximum achievable value of $\gamma$ can be estimated by equating
the synchrotron radiative looses to the power of the emitting
particles gained through the acceleration process, which in our case
can be written as:
\begin{equation}\label{}
    \frac{2}{3}\frac{e^{4}B^{2}\psi_{0}^{2}\gamma^{2}}{m^{2}c^{3}}=mc^{2}\gamma\Gamma_{LD}.
\end{equation}
Here
\begin{equation}\label{}
    \Gamma_{LD}=\frac{n_{b}\gamma_{b}\omega_{b}}{n_{p}\gamma_{p}^{5/2}},
\end{equation}
is the Landau damping rate, where $\omega_{b}=\sqrt{4\pi
e^{2}n_{b}/m}$ \citep{volo}. The estimations show, that the Lorentz
factor of the beam electrons that can be reached through the
acceleration process $\gamma\lesssim10^{19}$. Consequently, the
limit on the energy of synchrotron photons in our case should be
$\epsilon_{max}\sim10^{54}$eV (see Eq. (9)), which inevitably allows
generation of the observed $\sim100$GeV photons from the Crab pulsar
through the synchrotron emission mechanism.

\section{Spectrum of the synchrotron radiation}

To obtain the synchrotron emission spectrum in our case, we need to
solve the integral (8). For this reason, first let us find the
parallel distribution function of the beam electrons
$\textit{f}_{\parallel}$ during the QLD process of the cyclotron
instability. By multiplying both sides of Eq. (2) on $p_{\perp}$,
integrating it over $p_{\perp}$ and taking into account that the
distribution function vanishes at the boundaries of integration, Eq.
(2) reduces to
\begin{eqnarray}\label{16}
    \frac{\partial\textit{f}_{\parallel}}{\partial t}=\frac{\partial}{\partial
    p_{\parallel}}\left({\frac{\alpha_{s}}{m^{2}c^{2}\pi^{1/2}}p_{\perp_{0}}^{2}\textit{f}_{\parallel}}\right).
\end{eqnarray}

For $\gamma\psi\ll10^{10}$, a magnetic field inhomogeneity does not
affect the process of wave excitation. The equation that describes
the cyclotron noise level, in this case, has the form \citep{lomi}
\begin{equation}\label{18}
    \frac{\partial|E_{k}|^{2}}{\partial
    t}=2\Gamma_{c}|E_{k}|^{2},
\end{equation}
where
\begin{equation}\label{19}
   \Gamma_{c}=\frac{\pi^{2}e^{2}}{k_{\parallel}}\textit{f}_{\parallel}(p_{res}),
\end{equation}
is the growth rate of the instability. Here $k_{\parallel}$ can be
found from the resonance condition (1)
\begin{equation}\label{20}
     k_{\parallel_{res}}\approx\frac{\omega_{B}}{c\delta\gamma_{res}}.
\end{equation}
Combining Eqs. (10) and (11) one finds
\begin{equation}\label{21}
    \frac{\partial }{\partial t}\left\{\textit{f}_{\parallel}-\alpha\frac{\partial}{\partial
    p_{\parallel}}\left(\frac{|E_{k}|}{p_{\parallel}^{1/2}}\right)\right\}=0,
\end{equation}
\begin{equation}\label{22}
    \alpha=\left(\frac{4}{3}\frac{e^{2}}{\pi^{5}c^{5}}\frac{\omega_{B}^{6}\gamma_{p}^{3}}{\omega_{p}^{2}}\right)^{1/4}.
\end{equation}
Consequently, one can write
\begin{equation}\label{23}
    \left\{\textit{f}_{\parallel}-\alpha\frac{\partial}{\partial
    p_{\parallel}}\left(\frac{|E_{k}|}{p_{\parallel}^{1/2}}\right)\right\}=const.
\end{equation}
Taking into account that for the initial moment (the moment when the
cyclotron instability arises) the energy density of cyclotron waves
equals zero, the corresponding expression writes as
\begin{equation}\label{24}
   \textit{f}_{\parallel}-\alpha\frac{\partial}{\partial
    p_{\parallel}}\left(\frac{|E_{k}|}{p_{\parallel}^{1/2}}\right)=\textit{f}_{\parallel_{0}}.
\end{equation}
Let us assume that $|E_{k}|\propto\gamma^{-m}$ (as there is no
direct way to calculate the dependence $|E_{k}(\gamma)|$, we can
only make an assumption and check its plausibility by fitting the
theoretical emission spectrum with the observed one), in this case
for the parallel distribution function we will have
\begin{equation}\label{24}
   \textit{f}_{\parallel}\propto\textit{f}_{\parallel_{0}}+\gamma^{-m-\frac{3}{2}}.
\end{equation}
The initial distribution $\textit{f}_{\parallel_{0}}$ of the beam
particles in this case is the redistributed one after the Landau
damping process. The final shape of the distribution function of
resonant particles via the Landau damping when the stationary state
is reached is the plateau. But in our case this might not be reached
as in the same region develops the cyclotron instability. Thus we
consider $\textit{f}_{\parallel_{0}}\propto\gamma^{-n}$, where $n$
is not close to zero. Consequently, the parallel distribution of the
beam electrons is proportional to two power-law function with the
indexes $m+3/2$ and $n$.

The effective value of the pitch angle depends on $|E_{k}|^{2}$ as
follows \citep{ch10}
\begin{equation}\label{28}
    \psi_{0}=\frac{1}{2\omega_{B}}\left(\frac{3m^{2}c^{3}}{p_{\parallel}^{3}}\frac{\omega_{p}^{2}}{\gamma_{p}^{3}}|E_{k}|^{2}\right)^{1/4}.
\end{equation}
Using expressions (8), (18) and (19), and replacing the integration
variable $p_{\parallel}$ by $x=\epsilon/\epsilon_{m}$, we will get
the synchrotron emission spectrum
\begin{eqnarray}\label{29}
    F_{\epsilon}\propto\epsilon^{-\frac{2m+4n-1}{5-2m}}\left\{G_{1}\left(\frac{\epsilon}{\epsilon_{m}}\right)_{max}-G_{1}\left(\frac{\epsilon}{\epsilon_{m}}\right)_{min}\right\}+\nonumber\\
    +\epsilon^{-\frac{6m+5}{5-2m}}\left\{G_{2}\left(\frac{\epsilon}{\epsilon_{m}}\right)_{max}-G_{2}\left(\frac{\epsilon}{\epsilon_{m}}\right)_{min}\right\},
\end{eqnarray}
where
\begin{eqnarray}\label{29}
    G_{1}(y)=\int_{y}^{\infty} x^{\frac{2m+4n-1}{5-2m}}\left[\int_{x}^{\infty}K_{5/3}(z)dz \right] dx\nonumber
    \\
    G_{2}(y)=\int_{y}^{\infty} x^{\frac{6m+5}{5-2m}}\left[\int_{x}^{\infty}K_{5/3}(z)dz \right] dx.
\end{eqnarray}

The energy of the beam particles vary in a broad range
$\gamma_{b}\sim10^{6}-10^{9}$ in which case, we have
$(\epsilon/\epsilon_{m})_{max}\ll1$ and
$(\epsilon/\epsilon_{m})_{min}\gg1$. Under such conditions the
functions $G_{1}(y)\approx G_{2}(y)\approx G(0)$. Consequently, we
can assume that the synchrotron emission spectrum (Eq. (20))  is
proportional to two power-law functions
$\epsilon^{-\frac{2m+4n-1}{5-2m}}$ and
$\epsilon^{-\frac{6m+5}{5-2m}}$. According to VERITAS observations
the spectrum measured in the energy domain $(100-400)$GeV is well
described by power-law with the spectral index equal to $3.8$
\citep{veritas}. When $m\approx1$, we have $-(6m+5)/(5-2m)=-3.8$. At
the same time, when $m=1$ and $n=1.2$, we find
$-(2m+4n-1)/(5-2m)=-2$ that is in a good agreement with the
observations in the $10$MeV - $25$GeV energy domain, which shows the
power-law spectrum $F(\epsilon)\propto\epsilon^{-2.022\pm0.014}$
\citep{magic}.

\section{Discussion}

The interesting observational feature of the Crab pulsar of the
coincidence of pulse-phases from different frequency bands, ranging
from radio to VHE gamma-rays \citep{manch,magic, veritas} implies
that generation of these waves occur in one location of the pulsar
magnetosphere. This consideration automatically excludes the
generally accepted HE emission mechanisms, the Inverse Compton (IC)
scattering and the curvature radiation, which are not localized
\citep{difus,difus1}. These particular issues have been considered
by \citet{difus1}. Studying the curvature radiation it was found
that the curvature drift instability is efficient enough to rectify
the magnetic field lines (curvature tends to zero) in the region of
the generation of high and the VHE emission, making the curvature
emission process negligible. At the same time by analyzing the IC
scattering, it was found that for Crab pulsar's magnetospheric
parameters even very energetic electrons are unable to produce the
observed HE photons.

In \citet{lyu} it is argued that the main generation mechanism of
the VHE emission of the Crab pulsar is the IC scattering of the soft
UV photons by the secondary plasma particles. Particularly, it is
assumed that the secondary plasma, which is produced due to cascades
in the outer gaps of the magnetosphere is responsible for the soft
UV emission via the synchrotron mechanism. This radiation plays the
target field role for the IC scattering process. As a result, the
VHE $\gamma$-ray emission is produced extending to hundreds of GeV.
Let us consider the equation that gives the frequency of the photon
after the IC scattering (e.g. \citet{ry}):
\begin{equation}\label{}
    \omega'=\omega\frac{(1-\beta\cos\theta)}{1-\beta\cos\theta'+(1-\cos\theta'')\hbar\omega/\gamma
    mc^{2}},
\end{equation}
where $\omega$ is the frequency before scattering,
$\beta\equiv\upsilon/c$, $\theta=(\widehat{\textbf{PK}})$,
$\theta'=(\widehat{\textbf{PK}}')$ and
$\theta''=(\widehat{\textbf{KK}}')$ (by $\textbf{P}$ we denoted the
momentum of relativistic electrons before scattering, $\textbf{K}$
and $\textbf{K}'$ denotes the three momentum of photon before and
after scattering, respectively). The emission maximum comes along
the magnetic field lines. The pitch angles of the relativistic
electrons moving along the magnetic field lines are very small
($\psi\sim10^{-6}$ see Eq.(7)), accordingly as the UV emission is
generated through the synchrotron regime, the angle $\theta$ should
also be very small. Since we observe the well-localized pulses of
the VHE emission, the angle $\theta'\ll1$. Taking into account the
observational fact of the phase coincidence of signals from
different frequency bands, one should assume that
$\theta\approx\theta'$. At the same time the coincidence of pulse
signals of UV and the VHE emission gives $\theta''=0$. Consequently,
the IC scattering of the soft UV photons by the secondary plasma
electrons in case of the Crab pulsar should only cause the
redistribution of the electrons, but can not provide the increase in
the emission frequency, especially up to the VHE band. For
significantly increasing the photon energy by the IC scattering
processes without violating the condition of the pulse-phase
coincidence, the angle $\theta$ must be large enough, which can not
be provided by our model. Apparently, the IC scattering should play
the main role in the generation of the high energy emission for
pulsars with the large pulse profiles in soft energy domains.

The emission model proposed in the present paper implicitly explains
the observed pulse-phase coincidence of low (radio) and high
frequency (10MeV-400GeV) waves, as their generation is a
simultaneous process and it takes place in the same place of the
pulsar magnetosphere. In previous works \citep{difus1,ch10} we
applied this model to explain the HE (0.01-25GeV) pulsed emission of
the Crab pulsar observed by the MAGIC Cherenkov Telescope
\citep{magic}. It was found that on the light cylinder length-scales
the cyclotron instability is arisen, which on the quasi-linear stage
of the evolution causes re-creation of the pitch angles and as the
result the synchrotron radiation mechanism is switched on. We assume
that the source of the high and the VHE pulsed emission of the Crab
pulsar is the synchrotron radiation of the ultrarelativistic primary
beam electrons. To explain the observed high frequency gamma-rays by
synchrotron mechanism the Lorentz factors of the emitting particles
should be of the order of $10^{8}-10^{9}$ (see Eq. (9)). The highest
Lorentz factor for the typical pulsar is $\sim10^{7}$. Thus, we
assume that such an effective particle acceleration is caused by
existence of the Langmuir waves with the phase velocities
$\upsilon_{\varphi} \lesssim c$ close to the velocities of the
fastest beam electrons in the region close to the light cylinder.
Consequently, the L (electrostatic) waves are efficiently damped on
the most energetic primary beam electrons. The Landau damping causes
the growth of a HE tail on the distribution function of the resonant
electrons and inevitably throws the most energetic particles to
higher Lorentz factors (up to $\gamma\sim10^{9}$). At the same time
the beam electrons acquire the pitch angles due to the cyclotron
interaction with the transverse waves, which causes the synchrotron
radiation processes giving the observed high and the VHE emission up
to 400GeV. The distribution function tends to form the plateau (due
to Landau damping), though the number of processes impede this. The
reaction force of the synchrotron emission, scattering of L waves
and the Compton scattering of photons on the beam particles also
take place in process of formation of the distribution function of
beam electrons. As the result, it is unlikely for the distribution
to reach the shape of plateau and thus, we represent it as
$\textit{f}_{\parallel_{0}}\propto \gamma^{-n} $.

The calculation of the synchrotron emission spectrum by taking into
account the processes described above (see Eq. (20)) and matching it
with the observations shows that the emission spectrum in the
$10$MeV-$25$GeV energy domain depends on the power-law index $n$ of
distribution function of the beam particles. At the same time the
emission spectrum in $(100-400)$GeV energy domain does not depend on
$n$ but only depends on $m$ (see Eq. (18)). When $m=1$ and $n=1.2$
the emission spectrum well matches the measured one in both high
($0.01-25$GeV) and the very high ($100-400$GeV) energy domains. In
previous paper \citep{ch10} explaining the HE (10MeV-25GeV) spectrum
we obtained the power-law function with the exponential cutoff
$F_{\epsilon}\propto\epsilon^{-2}exp[-(\epsilon/23)^{1.6}]$, as
$\gamma_{b}\sim10^{8}$ were the highest considered Lorentz factors
of the emitting electrons. We assume that detection of the
exponential cutoff at higher ($>400GeV$) photon energies can not be
excluded, the exact location of the cutoff energy is defined by the
highest energy of the beam electrons that can be reached through the
Landau damping before the particles reach the light cylinder. This
particular problem needs more detailed investigation, which is the
topic of our future work.

\end{document}